\documentstyle[12pt]{article}
\def\lab{\label}

\def\eg{{\it e.g.\ }}

\def\lsim{\mbox{{\scriptsize \raisebox{-.9ex}
      {$\;\stackrel{{\textstyle <}}{\sim}\,$} }} }

\def\CPT{{\small $\chi$PT}}
\def\mN{m_{\mbox{\tiny N}}}

\begin{document}
\begin{titlepage}

\begin{flushright}
USC(NT)-0403
\end{flushright}

\vspace{1cm}
\begin{center}

{\Large\bf ASTROPHYSICAL WEAK-INTERACTION PROCESSES AND
NUCLEAR EFFECTIVE FIELD THEORY}\footnote{
Invited talk at the KIAS-APCTP Astro-Hadron Symposium,
KIAS, Seoul, Korea, November 10-14, 2003.}

\end{center}
\vspace{1cm plus 0.5cm minus 0.5cm}
\begin{center}
{\large 
K. Kubodera}\footnote{Work 
partially supported by the US 
National Science Foundation,
Grant No. PHY-0140214}

\end{center}
\vspace{0.5cm plus 0.5cm minus 0.5cm}
\begin{center}

{\it Department of Physics and Astronomy,\\
University of South Carolina, \\ 
Columbia, SC 29208, USA\\ 
E-mail: kubodera@sc.edu}

\end{center}
\vspace{1.0cm plus 0.5cm minus 0.5cm}

\begin{abstract}
Low-energy nuclear weak-interaction processes
play important roles in many astrophysical contexts,
and effective field theory is believed to be a highly 
useful framework for describing these processes 
in a model-independent manner. 
I present a brief account 
of the basic features of the nuclear effective theory
approach,
and some examples of actual calculations carried out 
in this method.
\end{abstract}
\end{titlepage}

\section{Introduction}

Low-energy nuclear weak-interaction processes 
play important roles
in many astrophysical phenomena and also 
in terrestrial experiments designed to detect
the astrophysical neutrinos. 
Obviously, it is important to have reliable estimates 
of the cross sections for these processes. 
I wish to describe here some of the recent developments
in our endeavor to obtain such estimates.\footnote{
This talk has some overlap with the one
I gave at NDM03~\cite{NDM03}.}
My main emphasis will be placed 
on comparison between the traditional method,
to be designated 
as the {\it standard nuclear physics approach}
(SNPA), and the newly developed 
{\it nuclear effective field theory} (EFT) approach.
I shall advocate the viewpoints that
(i) nuclear EFT can indeed be a powerful framework
for describing low-energy nuclear electroweak
processes and (ii) that, in practical applications, 
EFT and SNPA can play complementary roles. 

These points are nicely illustrated by  
the following three examples:
(i) neutrino-deuteron reactions for solar neutrino
energies;
(ii) solar pp fusion;
(iii) solar Hep fusion.
Since the process (iii) and related topics will
be discussed in detail by Dr. Tae-Sun Park 
at this Symposium, 
I shall concentrate on the first two reactions.
Let me start with a brief explanation of 
why these processes are 
of particular current interest.
 
At SNO (Sudbury Neutrino Observatory),
a one-kiloton heavy water Cerenkov counter
is used to detect the solar neutrinos.
SNO can monitor the neutrino-deuteron reactions:
\begin{eqnarray}
\nu_e+d &\rightarrow& 
 e^- + p + p\,,\,\,\,\,\,\,
\nu_x + d \rightarrow 
\nu_x+p + n\,, \nonumber\\
\bar{\nu}_e + d 
&\rightarrow& e^+ +n+n\,,\,\,\,\,\,
\bar{\nu}_x+d \rightarrow
\bar{\nu}_x+p+n\,,\label{nu-d}
\end{eqnarray}
as well as the pure leptonic reaction
$\nu_x+e^-\rightarrow \nu_x+e^-$.
Here $x$ stands for a neutrino 
of any flavor ($e$, $\mu$ or $\tau$).
The recent SNO experiments
\cite{ahmetal}
have established that
the total solar neutrino flux 
(counting all neutrino flavors) agrees with the prediction
of the standard solar model~\cite{bahcall},
whereas the electron neutrino flux 
from the sun is significantly smaller 
than the total solar neutrino flux.
The amount of deficit in the electron neutrino flux
was found to be consistent with what had been known as
the solar neutrino problem.
These results of the SNO experiments
have given clear evidence
for the transmutation of solar electron neutrinos into 
neutrinos of other flavors.
Obviously, a precise knowledge 
of the $\nu$-$d$ reaction cross sections 
is important
for the in-depth interpretation 
of the existing and future SNO data.

Meanwhile, 
the pp fusion reaction 
\begin{equation}
p+p\rightarrow d+e^++\nu_e \label{eq:pp}
\end{equation}
is the primary solar thermonuclear reaction
that essentially controls the luminosity of the sun,
and therefore the exact value of its cross section 
is a crucial input for 
any elaborate solar models.

\section{Calculational frameworks}

\subsection{Standard nuclear physics approach (SNPA)}

The phenomenological potential 
picture has been highly successful in describing 
a vast variety of nuclear phenomena.
In this picture an A-nucleon system
is described by a non-relativistic 
Hamiltonian of the form
\begin{equation}
H\,=\,\sum_i^A t_i + \sum_{i<j}^A V_{ij}
+\sum_{i<j<k}^A V_{ijk}+ \cdots\,, \label{HSNPA}
\end{equation} 
where $t_i$ is the kinetic energy of the $i$-th nucleon,
$V_{ij}$ is a phenomenological two-body potential
between the $i$-th and $j$-th nucleons,
$V_{ijk}$ is a phenomenological three-body potential, 
and so on.
(Since potentials involving three or more nucleons 
play much less important roles than
the two-body interactions, we shall be
concerned here mainly with $V_{ij}$.)
Once the model Hamiltonian $H$ is specified,
the nuclear wave function $|\Psi\!\!>$
is obtained by solving the Shr\"{o}dinger equation
\begin{equation}
H|\Psi\!>\,=\,E|\Psi\!>\,.\label{Sch}
\end{equation}
It is fortunate that the progress of 
numerical techniques for solving eq.(\ref{Sch}) has reached 
such a level \cite{cs98} that the wave functions 
of low-lying levels for light nuclei
can now be obtained with essentially no approximation
(once the validity of the model Hamiltonian 
eq.(\ref{HSNPA}) is accepted).
This liberates us from the 
``familar" nuclear physics complications 
that arise as a result of
truncating nuclear Hilbert space 
down to certain model space
(such as shell-model configurations within
a limited number of major shells, 
cluster-model trial functions, etc.)

We note that there is large freedom in selecting
possible forms of $V_{ij}$, 
apart from the well-established requirement that,
for a large enough value 
of the inter-nucleon distance,
$V_{ij}$ should agree with the
one-pion exchange Yukawa potential.
For the model-dependent short-range part 
of $V_{ij}$, the best we could do is to assume 
certain functional forms 
and fix the parameters contained in them
by demanding that the solutions of eq.(\ref{Sch}) 
for the A=2 case reproduce the nucleon-nucleon
scattering data 
(typically up to the pion-production threshold energy)
as well as the deuteron properties.
There are by now several so-called 
{\it modern high-precision}
{\it phenomenological} N-N potential
that can reproduce all the existing two-nucleon data
with normalized $\chi^2$ values close to 1.
These potentials differ widely
in the ways short-range physics is parametrized,
and, as a consequence,
they exhibit substantial difference
in their off-shell behaviors.
To what extent this arbitrariness may affect
the observables of our concern is an important
question, to which I will come back later.
  
In normal situataions,
nuclear responses
to external electroweak probes are 
given, to good approximation,  
by one-body terms, which are also called
the impulse approximation (IA) terms.
To obtain higher accuracy, however, 
we must include exchange current (EXC) 
terms,
which represent nuclear responses 
involving two or more nucleons.
These exchange currents 
(usually taken to be two-body operators)
are derived from one-boson exchange diagrams,
and the vertices featuring in the relevant diagrams
are determined to satisfy the low-energy theorems
and current algebra \cite{crit}.
We refer to a formalism based on this picture
as the {\it standard nuclear physics approach}
(SNPA). (This is also called a potential model
in the literature.)
Schematically, the nuclear matrix element in SNPA
is given by 
\begin{equation}
{\mathcal M}_{fi}^{\mbox{\tiny SNPA}}\,=\,
<\!\Psi_{\!f}^{\mbox{\tiny SNPA}}
\,|\sum_\ell^A{\mathcal O}_\ell^{\mbox{\tiny SNPA}}
+\sum_{\ell<m}^A 
{\mathcal O}_{\ell m}^{\mbox{\tiny SNPA}}
\,|\Psi_i^{\mbox{\tiny SNPA}}\!>\,, \label{ME-SNPA}
\end{equation}
where the initial (final) nuclear wave function,
$\Psi_i^{\mbox{\tiny SNPA}}$ 
($\Psi_f^{\mbox{\tiny SNPA}}$), 
is a solution of eq.(\ref{Sch});
${\mathcal O}_\ell^{\mbox{\tiny SNPA}}$ 
and ${\mathcal O}_{\ell m}^{\mbox{\tiny SNPA}}$ 
are, respectively,
the one-body and two-body transition operators for 
a given electroweak process.

SNPA has been used extensively 
to describe nuclear electroweak processes
in light nuclei, 
and generally good agreement
between theory and experiment~\cite{cs98}
gives a strong indication 
that SNPA essentially captures much 
of the physics involved.

\subsection{Effective field theory (EFT)}

Even though SNPA has been extremely successful
in correlating and explaining a wealth
of nuclear phenomana,
it is still important from a fundamental point of view
to raise the following issues.
First, since the hadrons and hadronic systems
(such as nuclei) are governed 
by quantum chromodynamics (QCD),
we should ultimately be able to relate 
nuclear phenomena with QCD,
but SNPA is reticent about this relation.
In particular, whereas chiral symmetry is known 
to be a fundamental symmetry of QCD,
the SNPA is largely disjoint from
this symmetry.
Second, even for describing low-energy phenomena,
SNPA starts with a ``realistic" phenomenological potential
which is tailored to encode short-range (high-momentum)
and long-range (low-momentum) physics
simultaneously.  This mixing of the two different scales
seems theoretically unsatisfactory. 
Third, as we write down a phenomenological Lagrangian
for describing the nuclear interaction and 
nuclear responses to the electroweak currents, 
SNPA does not offer us a clear guiding principle;  
it is not obvious whether SNPA is equipped with 
any identifiable expansion parameter
that helps us to control the possible forms of terms in 
the Lagrangian and that provides a general
measure of errors in our calculation.
To address these and other related issues,
a new approach based on EFT 
was proposed~\cite{wei90}
and it has been studied with great intensity;
for reviews, see Refs.~
\cite{bkm}$^-$\cite{br02}.
  
The intuitive picture of EFT is quite simple.
In describing phenomena
characterized by a typical energy-momentum scale $Q$,
we may expect 
that our Lagrangian need not contain explicitly
those degrees of freedom 
that belong to energy-momentum scales 
much higher than $Q$.
This expectation motivates us to
introduce a cut-off scale $\Lambda$
that is sufficiently large compared with $Q$
and classify our fields 
(to be generically represented by $\phi$)
into two groups: high-frequency fields 
$\phi_{\mbox{\tiny H}}$ 
whose frequencies are higher than $\Lambda$,
and low-frequency fields $\phi_{\mbox{\tiny L}}$
with frequencies lower than $\Lambda$.
By eliminating (or {\it integrating out})
$\phi_{\mbox{\tiny H}}$,
we arrive at an {\it effective} Lagrangian 
that only involves $\phi_{\mbox{\tiny L}}$
as explicit dynamical variables.  
In terms of path integrals,
the effective Lagrangian ${\mathcal L}_{{\rm eff}}$
is related to the original Lagrangian ${\mathcal L}$ as
\begin{eqnarray}
\int\![d\phi]
\exp\{{\rm i}\!\!\int \!\!d^4x{\mathcal L}[\phi]\}
&=&\int\![d\phi_{\mbox{\tiny H}}]
[d\phi_{\mbox{\tiny L}}]
\exp\{{\rm i}\!\!\int \!\!d^4x{\mathcal L}
[\phi_{\mbox{\tiny H}},\phi_{\mbox{\tiny L}}]\}\\
&\equiv& \int\![d\phi_{\mbox{\tiny L}}]
\exp\{{\rm i}\!\!\int\!\! d^4x{\mathcal L}_{\rm eff}
[\phi_{\mbox{\tiny L}}]\}\,.\label{EFTdef}
\end{eqnarray}

It can be shown that ${\mathcal L}_{{\rm eff}}$ 
defined by eq.(\ref{EFTdef})
inherits the symmetries 
(and the patterns of symmetry breaking, if there are any)
of the underlying Lagrangian ${\mathcal L}$.
It also follows that
${\mathcal L}_{{\rm eff}}$ should be
the sum of all possible monomials of 
$\phi_{\mbox{\tiny L}}$ and their derivatives
that are consistent with the symmetry requirements
of ${\mathcal L}$.
Because a term involving $n$ derivatives
scales like $(Q/\Lambda)^n$,
we can organize terms in ${\mathcal L}_{{\rm eff}}$ 
into a perturbative series 
in which $Q/\Lambda$ serves as an expansion parameter.
The coefficients of terms in 
this expansion scheme are called
the low-energy constants (LECs).
Provided all the LEC's up to a specified order $n$
can be fixed either from theory or from fitting 
to the experimental values of relevant observables,
${\mathcal L}_{{\rm eff}}$ serves as 
a complete (and hence model-independent) Lagrangian
to the given order of expansion.

Having sketched the basic idea of EFT,
we now discuss the specific aspects of EFT 
as applied to nuclear physics.
The underlying Lagrangian ${\mathcal L}$ 
in this case is the QCD Lagrangian ${\mathcal L}_{QCD}$,
whereas, for a typical nuclear physics
energy-momentum scale
$Q\ll \Lambda_{\chi}\sim 1$ GeV,
the effective degrees of freedom
that feature in ${\mathcal L}_{{\rm eff}}$ 
are the hadrons rather than the quarks and gluons.
It is non-trivial
to apply the formal definition in eq.(\ref{EFTdef})
to derive ${\mathcal L}_{{\rm eff}}$ 
written in terms of hadrons
starting from ${\mathcal L}_{QCD}$, because
the hadrons cannot be simply
identified with the low-frequency field 
$\phi_L$ in ${\mathcal L}_{QCD}$.
To proceed, we choose to be guided solely by 
symmetry considerations
and the above-mentioned expansion scheme. 
Chiral symmetry plays an important role here.  
Chiral symmetry is known to be spontaneously
broken, leading to the generation of the pions
as Nambu-Goldstone bosons.
We can incorporate this feature
by assigning suitable chiral transformation properties
to the Goldstone bosons
and writing down all possible chiral-invariant
terms up to a specified chiral order \cite{geo84}.
It is to be noted that the above consideration 
presupposes exact chiral symmetry in ${\mathcal L}_{QCD}$.
In reality, ${\mathcal L}_{QCD}$ contains
small but finite quark mass terms,
which violate chiral symmetry explicitly 
and lead to a non-vanishing value
of the pion mass $m_\pi$.
Again, there is a well-defined framework
to determine what terms are needed
to represent the effect of 
explicit chiral symmetry breaking
\cite{geo84}.
These considerations lead to an EFT 
called chiral perturbation theory 
($\chi$PT)~\cite{wei79,gl84}.
The successes of $\chi$PT in
the meson sector are well known;
see, {\it e.g.,} Ref.~\cite{bkm}. 

A difficulty we encounter in extending $\chi$PT
to the nucleon sector is that,
because the nucleon mass $\mN$ is comparable to 
the cut-off scale $\Lambda_{\chi}$, 
a simple application of expansion in $Q/\Lambda$
does not work.
We can surmount this obstacle with the use of 
heavy-baryon chiral perturbation theory (HB$\chi$PT),
which essentially consists in shifting 
the reference point of the nucleon energy 
from 0 to $\mN$ and integrating out 
the small component of the nucleon field 
as well as the anti-nucleon field.
Thus an effective Lagrangian in HB$\chi$PT 
contains as explicit degrees of freedom
the pions and the large components 
of the redefined nucleon field.
HB$\chi$PT has as expansion parameters
$Q/\Lambda_{\chi}$, $m_\pi/\Lambda_{\chi}$
and $Q/\mN$.
Since $\mN\approx \Lambda_{\chi}$,
it is convenient to combine chiral 
and heavy-baryon expansions
and introduce the chiral index ${\bar \nu}$ 
defined by $\bar{\nu}=d+(n/2)-2$.
Here $n$ is the number of fermion lines 
participating in a given vertex,
and $d$ is the number of derivatives
(with $m_\pi$ counted as one derivative). 
A similar power counting scheme can
be introduced
for Feynman diagrams as well.
According to Weinberg \cite{wei90},
a Feynman diagram that contains $N_A$ nucleons,
$N_E$ external fields,
$L$ loops and $N_C$ disjoint parts
scales like 
$(Q/\Lambda)^\nu$, where
the chiral index $\nu$ is defined as 
\begin{equation}
\nu = 2 L + 2 (N_C-1) + 2 - (N_A+N_E) + 
\sum_i \bar \nu_i\,,\lab{eq:nu}
\end{equation}
with the summation running over all the vertices.

Although HB$\chi$PT has been very successful
in the one-nucleon sector~\cite{bkm},
we cannot apply HB$\chi$PT 
in a straightforward manner 
to nuclei, which contain more than one nucleon.
This is because nuclei allow  
very low-lying excited states,
and the existence 
of this small energy scale
invalidates chiral counting~\cite{wei90}.
Weinberg avoided this difficulty by
classifying Feynman diagrams into two groups,
irreducible and reducible diagrams.
Irreducible diagrams are those
in which every intermediate state
has at least one meson in flight;
all others are categorized as reducible diagrams.
The chiral counting rules should only be applied
to irreducible diagrams.
The contribution of all the two-body irreducible diagrams 
(up to a specified chiral order)
is treated as an effective potential
(to be denoted by $V_{ij}^{\mbox{\tiny EFT}}$)
that acts on nuclear wave functions.
Meanwhile, the contributions of reducible diagrams
can be incorporated~\cite{wei90} 
by solving the Schr\"odinger equation
\begin{equation}
H^{\mbox{\tiny EFT}}
|\Psi^{\mbox{\tiny EFT}}\!>
\,=\,E|\Psi^{\mbox{\tiny EFT}}\!>\,,
\label{Sch-EFT}
\end{equation}
where
\begin{equation}
H^{\mbox{\tiny EFT}}
\,=\,\sum_i^A t_i + 
\sum_{i<j}^A V_{ij}^{\mbox{\tiny EFT}}\,, 
\label{HEFT}
\end{equation} 
We refer to this two-step procedure as
{\it nuclear} \CPT,
or, to be more specific,  
{\it nuclear} \CPT\  
in the Weinberg scheme.
(This is often called the $\Lambda$-counting scheme
\cite{lep99}.)

To apply nuclear \CPT\ to a process 
that involves (an) external current(s),
we derive a nuclear transition operator 
${\mathcal T}$ 
by calculating the contributions of 
all the irreducible diagrams
(up to a given chiral order $\nu$) 
that involve the relevant external current(s).
To maintain consistent chiral counting, 
the  nuclear matrix element of ${\mathcal T}$ 
must be calculated with the use of nuclear 
wave functions which are governed
by nuclear interactions that represent
all the irreducible A-nucleon diagrams 
up to $\nu$-th order. 
Thus, a transition matrix in nuclear EFT
is given by
\begin{equation}
{\mathcal M}_{fi}^{\mbox{\tiny EFT}}\,=\,
<\!\Psi_{\!f}^{\mbox{\tiny EFT}}
\,|\sum_\ell^A{\mathcal O}_\ell^{\mbox{\tiny EFT}}
+\sum_{\ell<m}^A
{\mathcal O}_{\ell m}^{\mbox{\tiny EFT}}
\,|\Psi_i^{\mbox{\tiny EFT}}
\!>\,, \label{ME-EFT}
\end{equation}
where the superscript, ``EFT", 
implies that the relevant quantities are obtained
according to EFT as described above.
If this program is carried out exactly, 
it would constitute an {\it ab initio} calculation.
It is worth noting that EFT tells us exactly
at what chiral order three-body operators 
start to contribute to ${\mathcal T}$, 
and that, to chiral orders of our present concern,
we do not need three-body operators.
For this reason 
we have retained in eq.(\ref{ME-EFT})
only one- and two-body operators. 
This type of unambiguous classification of 
transition operators according to their chiral orders
is a great advantage of EFT,
which is missing in eq.(\ref{ME-SNPA}).

I should mention that there exists
an alternative form of nuclear EFT
based on the power divergence subtraction (PDS) scheme. 
The PDS scheme proposed 
by Kaplan, Savage and Wise 
in their seminal papers~\cite{ksw}
uses a counting scheme (often called Q-counting)
that differs from the Weinberg scheme.
An advantage of the PDS scheme is
that it maintains formal chiral invariance,
whereas the Weinberg scheme loses
manifest chiral invariance.
In many practical applications, however,
this formal problem is not worrisome
up to the chiral order under consideration,
{\em i.e.,} the chiral order up to which 
our irreducible diagrams are evaluated.
Although the PDS scheme
has produced many important results
(for a review, see \eg 
Ref.~\cite{beaetal01}),
I concentrate here on the Weinberg scheme, 
because this is the framework in which
our own work has been done.

\subsection{Hybrid EFT}

In the above
I emphasized the formal merits of nuclear EFT.
In actual calculations, however, 
we face the following two problems.
First, it is still a great challenge
to generate, strictly within the EFT framework,
nuclear wave functions whose accuracy
is comparable to that of SNPA wave functions.
Second, as mentioned earlier,
the chiral Lagrangian, ${\mathcal L}_{{\rm eff}}$,
is definite only when the values of 
all the relevant LECs are fixed,
but there may be cases where this requirement
cannot be readily met.
A pragmatic solution to the first problem
is to use in eq.(\ref{ME-EFT}) wave functions
obtained in SNPA; 
we refer to this eclectic approach as hybrid EFT.
A nuclear transition matrix element in hybrid EFT
is given by 
\begin{equation}
{\mathcal M}_{fi}^{hyb-{\mbox{\tiny EFT}}}\,=\,
<\!\Psi_{\!f}^{\mbox{\tiny SNPA}}\,
|\sum_\ell^A{\mathcal O}_\ell^{\mbox{\tiny EFT}}
+\sum_{\ell<m}^A
{\mathcal O}_{\ell m}^{\mbox{\tiny EFT}}
\,|\Psi_i^{\mbox{\tiny SNPA}}\!>\,, \label{ME-hybrid}
\end{equation}
Because, as mentioned, the NN interactions
that generate SNPA wave functions
reproduce accurately the entirety of
the two-nucleon data, the adoption of eq.(\ref{ME-hybrid})
is almost equivalent to using the empirical data themselves
to control the initial and final nuclear wave functions.
In the context of theoretically deriving
the nuclear interactions based on EFT, 
hybrid EFT may look like ``retrogression".
But, if our goal is to obtain 
a transition matrix element as accurately as possible
with the maximum help of available empirical input,
hybrid EFT is a justifiable approach
insofar as the above-mentioned off-shell problem
and the contributions of three-body (and higher-body) 
interactions are properly addressed.  These points
will be discussed later on.

The calculations reported in Refs.~\cite{pkmr99,pc00}
seem to support hybrid EFT.
There, the nuclear matrix elements in the A=2 systems
for one-body operators (or IA terms)
calculated with the use of EFT-generated wave functions 
were found to be very close to those calculated with 
the SNPA wave functions.
Thus EFT and hybrid EFT should give practically
the same IA matrix elements.
Meanwhile, we can generally expect that the ratio of 
the two-body EXC contributions
to those of the IA operators should 
be much less sensitive to the details 
of the nuclear wave functions
than the absolute values are.
It therefore seems reasonable
to rely on \CPT\ for deriving transition operators
and evaluate their matrix elements
using the realistic wave functions obtained in SNPA,
and in this sense hybrid EFT is more than   
a mere expedient.

The issue of possible unknown LECs
will be discussed in the next subsection.

\subsection{MEEFT or EFT*}

Hybrid EFT can be used
for complex nuclei (A = 3, 4, ...) 
with essentially the same accuracy and ease
as for the A=2 system.
We should reemphasize in this connection
that, in A-nucleon systems (A$\ge$3),
the contributions of transition operators
involving three or more nucleons
are intrinsically suppressed 
according to chiral counting, 
and hence, up to a certain chiral order,
a transition operator in an A-nucleon system
consists of the same EFT-based
1-body and 2-body terms
as used for the two-nucleon system.
Then, since SNPA provides high-quality wave functions
for the A-nucleon system,
one can calculate 
${\mathcal M}_{fi}^{hyb-{\mbox{\tiny EFT}}}$
with precision comparable to that for 
the corresponding two-nucleon case.

Now, in most practical cases, the one-body
operator, ${\mathcal O}_\ell^{\mbox{\tiny EFT}}$, 
is free from unknown LECs.  So let us 
concentrate on the two-body operator, 
${\mathcal O}_{\ell m}^{\mbox{\tiny EFT}}$,
and suppose 
that ${\mathcal O}_{\ell m}^{\mbox{\tiny EFT}}$ 
under consideration contains an LEC (call it $\kappa$)
that cannot be determined with the use of 
A=2 data alone.
It is possible that 
an observable (call it $\Omega$)
in a A-body system (A$\ge$3)
is sensitive to $\kappa$ and 
that the experimental value of $\Omega$
is known with sufficient accuracy.
Then we can determine $\kappa$ 
by calculating 
${\mathcal M}_{fi}^{hyb-{\mbox{\tiny EFT}}}$
responsible for $\Omega$
and adjusting $\kappa$ to reproduce 
the empirical value of $\Omega$.
Once $\kappa$ is fixed this way,
we can make {\it predictions} for any other 
observables for any other nuclear systems
that are controlled by the same transition
operators.
When hybrid EFT is used in this manner,
we refer to it as MEEFT 
({\it more effective} EFT) or EFT*.

MEEFT is the most efficient existing formalism
for correlating various observables in different
nuclei, 
using the transition operators
controlled by EFT.
A further notable advantage of MEEFT 
is that, since correlating the observables 
in neighboring nuclei is likely to serve as 
an additional renormalization,
the possible effects of higher chiral order terms
and/or off-shell ambiguities 
can be significantly suppressed
by the use of MEEFT.\footnote{
MEEFT should be distinguished from
an earlier naive hybrid EFT model
in which the short-range terms were
dropped altogether using an intuitive argument
based on short-range NN repulsion.}
I will come back to this point later,
when I discuss concrete examples.  

We need to recall here 
the important role of momentum cutoff in EFT.
As emphasized before, the effective Lagrangian
${\mathcal L}_{eff}$ is, by construction, valid
only below the specified cutoff scale $\Lambda$.
This basic constraint must
be respected in our nuclear EFT calculations;
we must ensure that nuclear intermediate states 
involved in the computation of eq.(\ref{ME-EFT})
remain within this constrained regime.
It is reasonable to implement this constraint
by requiring that the two-nucleon relative momentum
should be smaller than $\Lambda$.
A possible choice
of the cutoff function is
the Gaussian form 
$\exp(-\vec{p}^2/\Lambda^2)$.
(The detailed form 
of the cutoff function should not be very important.) 
As a reasonable range of the value of $\Lambda$
we may choose: 
500 MeV $\lsim \Lambda \lsim$ 800 MeV,
where the lower bound is dictated by
the requirement that $\Lambda$
should be sufficiently large compared with  
the pion mass (in order to accommodate pion physics),
while the upper bound reflects the fact 
that our EFT is devoid of the $\rho$ meson.

\section{Numerical results}

We now discuss the applications
of the above-described calculational methods
to the two processes of our concern:
pp fusion and the $\nu$-$d$ reaction.
These reactions share the common feature that
a precise knowledge of the Gamow-Teller (GT) 
transition matrix elements is crucial
in estimating their cross sections.
We therefore concentrate on the GT transitions.
We will show here, following 
Refs.~\cite{PMS-pp,PMS-pphep},
that MEEFT can be used very profitably 
for these reactions.

We can argue (see, {\it e.g.,} Ref.~\cite{PMS-pphep})  
that 1-body IA operators for the GT transition
can be fixed unambiguously from the available
1-body data.
As for the 2-body operators, 
to next-to-next-to-next-to-leading order (N$^3$LO)
in chiral counting,
there appears one unknown LEC that cannot be
at present determined from data for the A=2 systems.
This unknown LEC, 
denoted by $\hat{d}_R$ in Ref.~\cite{pkmr98b},
parametrizes the strength
of contact-type four-nucleon coupling
to the axial current.
Park {\it et al.\ }\cite{PMS-pp,PMS-pphep}
noted that the same LEC, $\hat{d}_R$, 
also appears as a single unknown parameter
in the calculation of the tritium $\beta$-decay rate
$\Gamma_{\beta}^t$,
and they used MEEFT
to place a constraint on $\hat{d}_R$
from the experimental value of $\Gamma_{\beta}^t$.
Since $\Gamma_{\beta}^t$(exp) 
is known with high precision,
and since the accurate wave functions of 
$^3$H and $^3$He are available
from a well-developed 
SNPA calculation~\cite{Metal},
we can determine $\hat{d}_R$
with sufficient accuracy for our purposes.
Once the value of $\hat{d}_R$ is determined this way, 
we can carry out parameter-free MEEFT calculations 
for pp-fusion~\cite{PMS-pp,PMS-pphep}
and the $\nu$-$d$ reactions~\cite{andetal-nud}.
I present here a brief summary of the results of
these calculations.  

For a given value of $\Lambda$
within the above-mentioned range
(500 MeV $\lsim \Lambda \lsim$ 800 MeV),
$\hat{d}_R$ is adjusted to reproduce 
$\Gamma_{\beta}^t$(exp),
and then the cross sections for pp-fusion
and the $\nu d$ reactions are calculated.
The results indicate 
that, although the best-fit value of $\hat{d}_R$ 
varies significantly as a function of $\Lambda$,
the observables (in our case the above two reaction
cross sections) exhibit a high degree of stability
against the variation of $\Lambda$.
This stability may be taken as an indication
that the use of MEEFT for
inter-correlating the observables in neighboring nuclei
{\it effectively} renormalizes various effects, 
such as the contributions of higher-chiral order terms,
mismatch between the SNPA and EFT wave functions,
etc.
This stability is essential in order 
for MEEFT to maintain its predictive power. 

Park {\it et al.}~\cite{PMS-pp,PMS-pphep}
used MEEFT
to calculate the rate of pp fusion,
$pp\!\rightarrow\!e^+\nu_e d$.
The result expressed in terms of
the threshold $S$-factor is
\begin{equation}
S_{pp}(0)=3.94\!\times\!(1\pm0.005)
\times 10^{-25}\,{\rm MeV\, b}\,.\label{Spp}
\end{equation}
It has been found that $S_{pp}(0)$ changes 
only by $\sim$0.1\%
against changes in $\Lambda$,
assuring thereby the robustness
of the MEEFT prediction.
The MEEFT result, eq.(\ref{Spp}),
is consistent with that obtained in SNPA by Schiavilla 
{\it et al.}~\cite{schetal98}. 
Meanwhile, the fact that MEEFT allows us 
to make an error estimate [as given in 
eq.(\ref{Spp})] is a notable advantage
over SNPA.
The details on how we arrive at this error estimate
can be found in Refs.~\cite{PMS-pp,PMS-pphep}.
Here I just remark that the error 
indicated in eq.(\ref{Spp})
represents an improvement 
by a factor of $\sim$10
over the previous results
based on a naive hybrid EFT~\cite{pkmr98b}.

We now move to the $\nu$-$d$ reactions, eq.(\ref{nu-d}),
and give a brief survey 
of all the recent results obtained in SNPA, EFT
and MEEFT.
Within SNPA a detailed calculation of 
the $\nu$-$d$ cross sections, $\sigma(\nu d)$,
was carried out by Nakamura, Sato,
Gudkov and myself~\cite{NSGK},\footnote{
For a review of the earlier SNPA calculations,
see Ref.~\cite{kn94}.}
and this calculation has recently been updated
by Nakamura {\it et al.}\ (NETAL)~\cite{NETAL}.
As demonstrated in Ref.\cite{crsw},
the SNPA exchange currents for the GT transition
are dominated 
by the $\Delta$-particle excitation diagram,
and the reliability of estimation
of this diagram depends on the precision
with which the coupling constant 
$g_{\pi N\Delta}$ is known.
NETAL fixed $g_{\pi N\Delta}$
by fitting $\Gamma_{\beta}^t$(exp),
and proceeded to calculate $\sigma(\nu d)$.
Meanwhile, Butler, Chen and Kong (BCK)~\cite{EFT} 
carried out
an EFT calculation of the $\nu$-$d$ cross sections,
using the PDS scheme~\cite{ksw}.
The results of BCK
agree with those of NETAL in 
the following sense.
BCK's calculation involves one unknown LEC 
(denoted by $L_{\rm 1A}$), 
which like $\hat{d}_R$ in Ref.\cite{PMS-pphep},
represents the strength of  
a four-nucleon axial-current coupling term.
BCK determined $L_{\rm 1A}$
by requiring that
the $\nu d$ cross sections of NETAL be reproduced
by their EFT calculation.
With the value of $L_{\rm 1A}$
adjusted this way,
$\sigma(\nu d)$'s obtained by BCK
show a perfect agreement with
those of NETAL for all the four reactions 
in eq.(\ref{nu-d}) and
for the entire solar neutrino energy range,
$E_\nu\lsim$ 20 MeV.
Moreover, the best-fit value,
$L_{\rm 1A}=5.6\,{\rm fm}^3$, found 
by BCK~\cite{EFT} is consistent 
with its magnitude
expected from the naturalness argument 
(based on a dimensional analysis), 
$|L_{\rm 1A}|\le 6\,{\rm fm}^3$.
The fact that an EFT calculation
(with one parameter fine-tuned)
reproduces the results of SNPA very well
strongly suggests the robustness of
the SNPA calculation of $\sigma(\nu d)$.

Even though it is reassuring 
that the $\nu$-$d$ cross sections
calculated in SNPA and EFT
agree with each other 
(in the above-explained sense),
it is desirable to carry out an EFT calculation 
that is free from  any adjustable LEC.
Fortunately, MEEFT allows us
to carry out an EFT-controlled parameter-free calculation
of the $\nu$-$d$ cross sections,
and such a calculation was carried out 
by Ando {\it et al.}~\cite{andetal-nud}.
The $\sigma(\nu d)$'s obtained in Ref.~\cite{andetal-nud}
are found to agree within 1\% with
$\sigma(\nu d)$'s obtained by NETAL 
using SNPA~\cite{NETAL}. 
These results show 
that the $\nu$-$d$ cross sections used 
in interpreting the SNO experiments~\cite{ahmetal}
are reliable at the 1\% precision level.

We remark that, as PDS~\cite{ksw}
is built on an expansion scheme 
for transition amplitudes themselves,
it does not employ the concept of wave functions.
This feature is an advantage
in some contexts,
but its disadvantage in the present context
is that we cannot readily relate 
the transition matrix elements
for an A-nucleon system with those for 
the neighboring nuclei;
in PDS, each nuclear system requires a separate 
parametrization.
This feature underlies the fact
that, in the work of BCK~\cite{EFT},
$L_{1A}$ remained undetermined,
because no experimental data is available 
to fix $L_{1A}$
within the two-nucleon systems.

\section{Discussion}

In introducing hybrid EFT, we replace 
$|\Psi^{\mbox{\tiny EFT}}\!>$ for the initial
and final nuclear states in eq.(\ref{ME-EFT})
with the corresponding $|\Psi^{\mbox{\tiny SNPA}}\!>$'s;
see eq.(\ref{ME-hybrid}).
This replacement may bring in
a certain degree of model dependence,
called the off-shell effect, because
the phenomenological NN interactions 
are constrained only by the 
on-shell two-nucleon observables.\footnote{
In a fully consistent theory, physical observables
are independent of field transformations
that lead to different off-shell behaviors, 
and therefore the so-called off-shell effect 
is not really a physical effect.
In an approximate theory, observables may 
exhibit superficial dependence on the off-shell behavior,
and it is customary to refer to this dependence
as an off-shell ``effect".} 
This off-shell effect, however, is expected 
to be small for the reactions under consideration,   
since they involve low momentum transfers
and hence are not extremely sensitive to
the short-range behavior of the nuclear wave functions.
One way to quantify this expectation is to compare
a two-nucleon relative wave function generated 
by the phenomenological
potential with that generated by an EFT-motivated potential.
Phillips and Cohen~\cite{pc00} made such a comparison
in their analysis of the 1-body operators 
responsible for electron-deuteron Compton scattering,
and showed that a hybrid EFT works well up to
momentum transfer 700 MeV.  
A similar conclusion is expected to hold
for a two-body operator,
so long as its radial dependence
has a ``smeared-out" structure reflecting 
the finite momentum cutoff. 
We can therefore expect that
hybrid EFT as applied to low energy
should be practically
free from the off-shell ambiguities.
The off-shell effect should be even less significant 
in MEEFT, wherein an additional ``effective" renormalization
is likely to be at work (see subsection 2.4).
 
Another indication of the stability 
of the MEEFT results comes from
a recently proposed idea of the 
low-momentum nuclear potential~\cite{kuo}.
As mentioned, a ``realistic phenomenological" nuclear 
interaction, $V_{ij}$ in eq.(\ref{HSNPA}),
is determined by fitting 
to the full set of two-nucleon data 
up to the pion production threshold energy. 
So, physically, $V_{ij}$ should reside in a momentum regime
below a certain cutoff, $\Lambda_c$.
In the conventional treatment, however, 
the existence of this cutoff scale is ignored,
and eq.(\ref{Sch}) is solved in such a manner that
the entire momentum range is allowed to participate.  
Bogner {\it et al.\ } proposed to construct an 
{\it effective low-momentum} potential 
$V_{low-k}$ by eliminating 
(or integrating out) from $V_{ij}$
the momentum components higher than
$\Lambda_c$, and calculated $V_{low-k}$'s 
corresponding to a number of well-established
of $V_{ij}$'s.  
It was found that
all these $V_{low-k}$'s lead to
identical half-off-shell T-matrices,
even though the ways short-range physics is encoded
in them are highly diverse.
This implies that the $V_{low-k}$'s are free from
the off-shell ambiguities, and therefore the use of 
$V_{low-k}$'s is essentially equivalent to employing 
$V_{ij}^{\mbox{\tiny EFT}}$ 
(appearing in eq.(\ref{HEFT})),
which by construction should be model-independent.
Now, as mentioned, our MEEFT calculation has
a momentum-cutoff regulator built in, 
and this essentially
ensures that the matrix element,
${\mathcal M}_{fi}^{hyb-{\mbox{\tiny EFT}}}$, 
in eq.(\ref{ME-hybrid})
is only sensitive to the half-off-shell T-matrices
that are controlled by $V_{low-k}$ instead of $V_{ij}$.
Therefore, we can expect that the MEEFT results 
reported here are essentially free from the off-shell
ambiguities.

\section{Summary}

After giving a very limited survey
of nuclear \CPT,
I must repeat my disclaimer
that I have left out many important topics
belonging to nuclear \CPT.
Among others,
I did not discuss very important studies 
by Epelbaum, Gl\"{o}ckle and Mei{\ss}ner~\cite{epeetal}
to construct a formally consistent framework
for applying \CPT\ to complex nuclei.  
It should be highly informative
to apply this type of formalism
to electroweak processes and compare 
the results with those of MEEFT.
In this connection I find it noteworthy
that the range of the cutoff parameter
favored in Ref.~\cite{epeetal}
is consistent with the range used 
by Park {\it et al.}~\cite{PMS-pp,PMS-pphep}

Despite the highly limited scope of topics covered,
I hope I have succeeded in demonstrating  
that MEEFT is a powerful framework for 
computing the transition amplitudes of
low-energy electroweak processes in light nuclei.
I also wish to emphasize that,
in each of the cases for which 
both SNPA and MEEFT calculations
have been performed, it has been found that 
the result of MEEFT supports
and improves the SNPA result.  

\section*{Acknowledgments}
This talk is based on the work done 
in collaboration with T.-S. Park, M. Rho, 
D.-P. Min, F. Myhrer, T. Sato,
V. Gudkov and S. Nakamura,
and I wish to express my sincere thanks 
to these colleagues.

\end{document}